\documentclass[runningheads]{llncs}

\usepackage{tipa}
\usepackage{booktabs}
\usepackage{xspace}
\usepackage{graphics}
\usepackage{todonotes}
\usepackage{xcolor}
\usepackage{hyperref}
\usepackage[para]{footmisc}

\definecolor{mygray}{gray}{0.45}
\newcommand{\light}[1]{\textcolor{mygray}{#1}}

\newcommand{\io}[1]{#1}
\newcommand{\cm}[1]{#1}
\newcommand{\sm}[1]{#1}
\newcommand{\edit}[1]{#1}

\usepackage[numbers,sort&compress]{natbib}

\usepackage{url}

\begin{document}
\mainmatter              %
\title{Reproducing Personalised Session Search \\ over the AOL Query Log}
\author{Sean MacAvaney \and Craig Macdonald
\and Iadh Ounis}
\authorrunning{MacAvaney et al.} %
\institute{University of Glasgow, United Kingdom\\
\email{\{first.last\}@glasgow.ac.uk}}

\maketitle              %

\begin{abstract}
Despite its troubled past, the AOL Query Log continues to be an important resource to the research community---particularly for tasks like search personalisation. When using the query log \io{these} ranking \io{experiments}, \edit{little attention is usually paid to the document corpus}. Recent work typically uses a corpus containing versions of the documents collected long after the log was produced. Given that web documents are prone to change over time, we study the differences present between a version of the corpus containing documents as they appeared in 2017 (which has been used by several recent works) and a new version we construct that includes documents close to as they appeared at the time the query log was produced (2006). We demonstrate that this new version of the corpus has a far higher coverage of documents present in the original log (93\%) than the 2017 version (55\%). Among the overlapping documents, the content often differs substantially. Given these differences, we re-conduct session search experiments that originally used the 2017 corpus and find that \sm{when using our corpus for training or evaluation, system performance improves. We place the results in context by introducing recent adhoc ranking baselines. We also confirm the navigational nature of the queries in the AOL corpus by showing that including the URL substantially improves performance across a variety of models.} Our version of the corpus can be easily reconstructed by other researchers and is included in the \texttt{ir-datasets} package.
\end{abstract}
\section{Introduction}

When released in 2006, the AOL Query Log~\cite{pass2006picture} drew harsh criticism from the media over privacy concerns~\cite{barbaro_zeller_2006}. Since then, however, it has been an important resource to the research community (e.g.,~\cite{Nunes2008UseOT,Rafiei2010DiversifyingWS}). Even to this day, the AOL Query Log continues to enable studies in analysis of data leaks~\cite{Kamara2021CryptanalysisOE}, search autocompletion~\cite{Kang2021QueryBlazerEQ}, weak supervision for adhoc search~\cite{Dehghani2017NeuralRM,MacAvaney2019ContentBasedWS}, search result personalisation~\cite{Huang2021TPRMAT,Ma2020PSTIETI}, and session-based search~\cite{Ahmad2018MultiTaskLF,Ahmad2019ContextAD,Qu2020ContextualRW,Zhu2021ContrastiveLO,Cheng2021LongSS}.

A key limitation of the AOL Query Log is that it does not include document contents; it only provides a user identifier, query text, query date/time, and the URL and rank of clicked documents (if any). This means that for studies that use the logs as a training and benchmark data for tasks like search result personalisation and session search, a document corpus needs to be constructed. \sm{Often the approach used for constructing the corpus is unspecified. To the best of our knowledge, when the approach is specified, it always involves scraping \textit{current} versions of the documents.} There are two main problems with this approach. First, given that the contents of web documents are highly prone to change over time, recent versions of the documents may not reflect the contents of the documents as they appeared to the users. Second, this approach impedes reproduciblity and replicability efforts in the area, since the contents of the documents cannot be released publicly due to the potential that they contain copyrighted material.

In this paper, we study the effect that the document corpus used for AOL Query Log experiments has on reproducibility. We start by building a new document corpus that attempts to better reflect the documents present in the AOL Query Log as they appeared when the log was collected. This is accomplished by using the Internet Archive,\footnote{\url{https://archive.org/}} and thus we refer to our corpus as \texttt{AOLIA}. We find that this approach is able to cover far more of the documents that appeared in the AOL Query log (93\%) when compared to a commonly-used version of the corpus that was collected in 2017 (55\%, shared on request by \citet{Ahmad2019ContextAD}). Based on the timestamps from the Internet Archive, we are confident that the documents in \texttt{AOLIA} also better reflect the content of the documents as they appeared at the time, with 86\% of the documents coming from during or in the three months prior to the log. We find that the content of the overlapping documents changed substantially in the 11-year period, with 28\% of documents having no token overlap in the title \edit{(which is often used for session-based search~\cite{Ahmad2019ContextAD,Qu2020ContextualRW,Huang2021TPRMAT})}.

\sm{\edit{We further conduct a reproducibility and replicability study\footnote{\edit{ACM version 1.1 definitions of reproducibility and replicability: \url{https://www.acm.org/publications/policies/artifact-review-and-badging-current}.}}} of personalised session search tasks based on the AOL Query Log. \edit{We are unable to \textit{reproduce} results using the 2017 version of the corpus, but our \textit{replication} results (using \texttt{AOLIA}) are more in line with the original findings.} To put the results in context, we also include a neural adhoc ranking baseline, which ultimately outperforms the methods we investigate. We also study the effect of using the document's URL as additional text and find that it improves the performance of all methods we \edit{investigate} (often by a large margin), further confirming the navigational nature of the queries in the AOL Query Log. In summary, our contributions are:}

\begin{enumerate}
\item We provide an alternative document corpus (\texttt{AOLIA}) for the AOL Query Log based on versions of the documents as they were likely to have appeared at the time the query log was collected.
\item We release artifacts and software such that other researchers will be able to construct \texttt{AOLIA} themselves, \sm{promoting reproducibility}.
\item \sm{We study the reproducibility \edit{and replicability} of three session-based search approaches, and find that using \texttt{AOLIA} alone can improve the performance of session-based search systems due to higher-quality documents, and that \io{the} training and evaluation datasets constructed from \texttt{AOLIA} can be considerably larger due to the increased coverage of the dataset.}
\end{enumerate}

\sm{The remainder of this paper is organised as follows. In Section~\ref{sec:back} we provide additional background information about the problem. Then, Section~\ref{sec:constr} details our process for constructing \texttt{AOLIA}. Section~\ref{sec:comp} provides a comparison between \texttt{AOLIA} and a version produced in 2017 that is used by several recent works. Section~\ref{sec:repr} then focuses on reproducing prior works using \texttt{AOLIA}. Finally, Section~\ref{sec:limit} details the limitations of our approach, and Section~\ref{sec:conc} draws final conclusions.}

\section{Background}\label{sec:back}

Past works that \edit{make} use of the AOL Query Log \edit{use} recent versions of the log's clicked documents. Because the content of web pages can change over time, using recent versions necessitates \io{a filtering process, which removes} query-document pairs that are no longer relevant. For instance, \citet{Ahmad2019ContextAD} reports \textit{``...in our preliminary experiments, we observed that many recorded clicks do not have lexical overlap concerning the queries. One possible reason is that we crawled the recorded clicks from the AOL search log in 2017 and many of the clicked documents’ content updated since 2006 when the AOL log was recorded.''}

\edit{Several} alternatives to The Internet Archive \edit{exist} as sources of data for a reproducible AOL corpus. Although the Common Crawl\footnote{\url{https://commoncrawl.org/}} would provide a more comprehensive corpus (i.e., it includes a more natural selection of documents, rather than only documents clicked by the user), we show in Section~\ref{sec:constr} that the AOL corpus at the time likely did not contain a representative sample of documents from the web, but rather focused heavily on home pages. \io{Moreover}, since the oldest version of the Common Crawl is from 2008--09, the content of the documents may already have changed since the time of the log. Finally, the size of the relevant archives (hundreds of terabytes) could add substantial difficulty in downloading and working with the data. The ClueWeb 2009\footnote{\url{https://lemurproject.org/clueweb09/}} and 2012\footnote{\url{https://lemurproject.org/clueweb12/}} corpora would be another option, and are appealing given that many research groups already have a copy of them. However, like the Common Crawl, they reflect the contents of documents several years after the log was constructed. \edit{Furthermore, there is} low coverage of the target URLs in the ClueWeb corpora.

\sm{Other efforts \edit{investigate} the stability of using mutable web resources as document corpora in IR. \citet{McCreadie2012OnBA} \edit{find} that naturally-occurring deletions from the Twitter corpus used by the TREC Microblog tasks do  not have a substantial \io{effect} on the results of experiments that use the corpus. However, the situation for general web pages is different because the content can change over time (tweets can only be deleted, not updated). Despite these findings,}
\citet{Sequiera2017FinallyAD} \edit{investigate} the use of the Internet Archive as an alternative source of data for the TREC 2013--14 Microblog corpus. Our work not only differs in terms of the document corpus targeted, but also the download mechanism; the Twitter stream they \edit{use} is conveniently bundled by month by the Internet Archive, whereas there is no such bundle available for the documents present in the AOL Query Log. \sm{Consequently, the steps involved to build our version of the AOL corpus \edit{are} necessarily more complicated.
}

\section{Reconstructing the AOL Document Corpus}\label{sec:constr}

In this section, we reconstruct a document corpus that better reflects the documents as they appeared to the users at the time. Through this process, we create artifacts and software that others can use to construct this dataset themselves, further \sm{promoting} reproducibility in this area.

We start by building a set of all unique URLs that appear in the AOL Query Log.\footnote{\url{http://www.cim.mcgill.ca/~dudek/206/Logs/AOL-user-ct-collection/aol-data.tar.gz}} Importantly, we acknowledge that this only represents documents that users clicked; the full list of documents indexed by AOL at the time is not available. This \edit{process} \sm{results in} 1,632,620 unique URLs. Nearly half (48.7\%) of \io{the} URLs were only clicked a single time. All but 15 URLs specify either \texttt{http} or \texttt{https} URI schemes (14  specify \texttt{ftp} and one specifies \texttt{about}). 98.4\% of the URLs refer to the home page of a website (i.e., have no path), which suggests that the search engine primarily functioned as a navigational tool at the time.

\begin{figure}[t]
\centering
\includegraphics[scale=0.7]{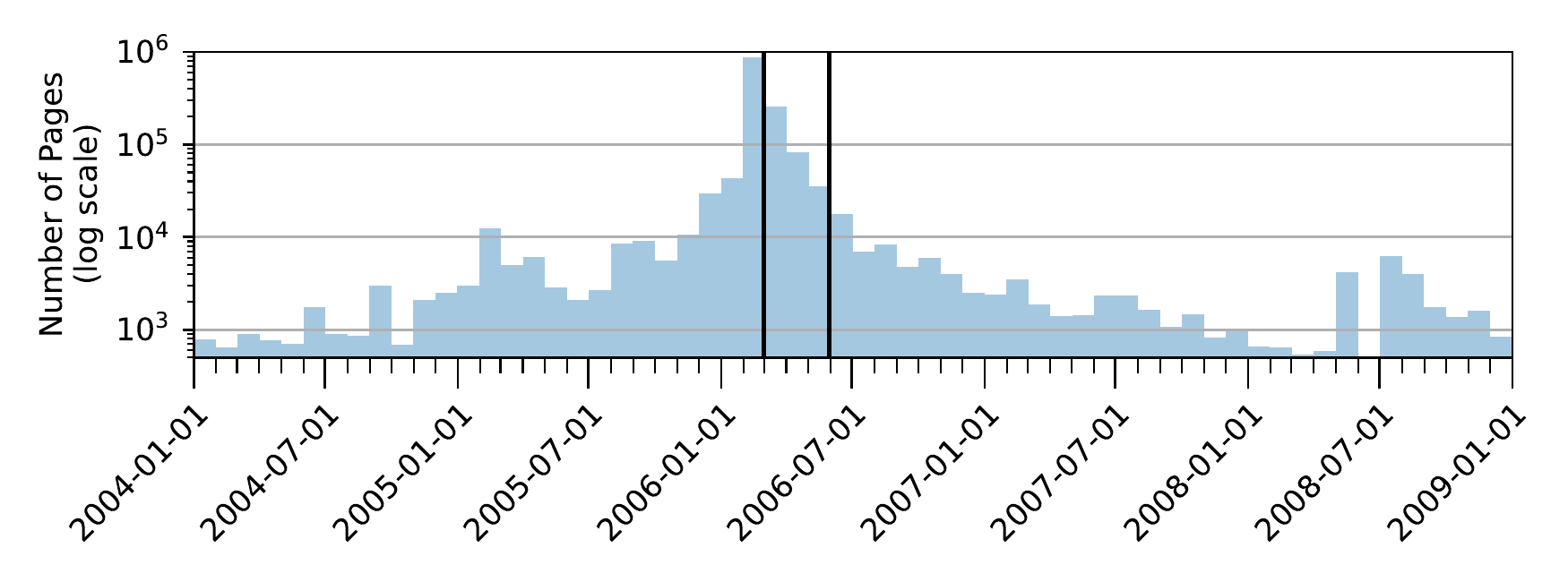}
\caption{Distribution of the archive dates of the web pages in the scraped collection from 2004--2008. The vast majority of pages \edit{are} recovered from the time period during or right before the AOL Query Log (marked by vertical lines).}
\label{fig:dates}
\end{figure}

We then query archive.org's WayBack Machine's availability API\footnote{\edit{API Endpoint:} \url{https://archive.org/wayback/available}} to request a version of the page as it appeared as close as possible to 1 March 2006 (the beginning of the AOL Query Log). Remarkably, we \sm{find} that 93\% of URLs were archived. Fig.~\ref{fig:dates} shows the distribution of the dates of the archived pages. The vast majority of the found URLs (84\%) \sm{are} from the period during or in the three months prior to the log (January to May 2006). A further 8\% \sm{are} from before 2006, and a total of 96\% of pages \sm{have} an archived copy before 2007. Based on these dates, we feel that the corpus represents a reasonable approximation of \io{the} \sm{documents present in the query log at the time it was collected.}

We then fetch the archived versions of the documents and parse the resulting HTML using a libxml2-based parser. The title and body text are extracted (discarding content that appears in non-content tags, such as \texttt{<script>}). A small number of documents (0.3\%) encountered either parsing errors or persistent \edit{download errors}. \edit{We discard these documents}. The median title length is 5 tokens (interquartile: (3, 9)), while the median body length is 198 tokens (interquartile: (47, 454)). As is the case for web content, some documents are substantially longer (up to 1.7M tokens). When compressed, the corpus is 3.4G in size. Using a FastText~\cite{joulin2016bag} language classifier\footnote{\url{https://fasttext.cc/docs/en/language-identification.html}} over the document title and body, we find that the vast majority of documents (92.5\%) likely contain English text, as expected. Table~\ref{tab:langs} presents a further breakdown of the top languages in \texttt{AOLIA}. The breakdown is similar to that of the queries that appear in the log, \sm{when considering \io{that}} language identification is more prone to errors for short texts like keyword queries.

\begin{table}[t]
\centering
\caption{Top languages present in the \texttt{AOLIA} corpus, compared to the prevalence of \edit{the} language of queries in the log.}
\label{tab:langs}
\begin{tabular}{lrr|lrr}
\toprule
Language & Corpus & Queries & Language & Corpus & Queries \\
\midrule
English &  92.5 \% & 79.6 \% & Japanese   & 0.4 \% & 0.1 \% \\
French &    1.6 \% &  3.2 \% & Portuguese & 0.3 \% & 0.8 \% \\
Spanish &   1.4 \% &  2.1 \% & Dutch      & 0.3 \% & 0.9 \% \\
German &    1.1 \% &  2.6 \% & Russian    & 0.2 \% & 0.4 \% \\
Italian &   0.5 \% &  1.8 \% & All Others & 1.5 \% & 8.5 \% \\
\bottomrule
\end{tabular}
\end{table}

Though we cannot distribute the contents of this corpus directly due to potentially copyrighted content, we take the following steps to facilitate reproducibility using this dataset:

\begin{enumerate}
\item We publicly release a mapping of the Internet Archive URLs so that others can fetch the same versions of the original documents.\footnote{\url{https://macavaney.us/aol.id2wb.tsv.gz}}
\item We provide software to download and extract the contents of these documents.\footnote{\url{https://github.com/terrierteam/aolia-tools}}
\item We include a new \texttt{aol-ia} dataset in the \texttt{ir-datasets}~\cite{macavaney:sigir2021-irds} package, \io{which} provides easy access to this document corpus and the AOL log records. The package automatically downloads the log records from a public source, the Internet Archive mapping (from 1), and provides instructions to the user on how to run the extraction software (from 2). Once built, the dataset can easily be used by tools like PyTerrier~\cite{macdonald:cikm2021-pyterrier} and OpenNIR~\cite{macavaney:wsdm2020-onir}.
\end{enumerate}

\section{Comparing \texttt{AOLIA} with \texttt{AOL17}}\label{sec:comp}

\sm{In this section, we compare \texttt{AOLIA} with a version of the corpus that used more recent versions of the documents present in the log. Specifically, we use the version first used by \citet{Ahmad2018MultiTaskLF}, which uses documents from the AOL Query Log as they appeared in 2017 (so we call this corpus \texttt{AOL17}). This corpus has been used by other works (e.g.,~\cite{Ahmad2019ContextAD,Qu2020ContextualRW,Huang2021TPRMAT}).} We treat \texttt{AOL17} as a representative example of a contemporary version of the AOL corpus\sm{, noting that more recent versions of the corpus are likely to diverge even further from the original documents.}

\looseness -1 Table~\ref{tab:ds_comparision} provides a comparison between the URLs present in the two datasets. In terms of absolute coverage, \texttt{AOLIA} provides a high (albeit still incomplete) coverage of 93\%. Meanwhile, \texttt{AOL17} contains only 55\% of the URLs found in the log. There are roughly 12k URLs found in \texttt{AOL17} but not in \texttt{AOLIA}. \sm{\textit{Content pages} (i.e., non-homepages) \sm{are} over-represented among these documents, constituting 13\% (1,637) of pages (compared to 2\% of the overall corpus).} \texttt{AOLIA} compensates for this disparity simply by virtue of size, filling in 639,747 documents missing from \texttt{AOL17} (9,803 of which are content pages). Even though adding missing documents to \texttt{AOLIA} from \texttt{AOL17} would increase the total coverage from 93.4\% to 94.2\%, doing so would \cm{reduce} reproducibility, since those documents may contain copyrighted material and therefore cannot be distributed publicly.

\begin{table}[t]
\centering
\caption{Comparisons of URLs present in the \texttt{AOLIA} and \texttt{AOL17} datasets. The Total column indicates the percentage of all URLs present in the AOL corpus.}
\label{tab:ds_comparision}
\begin{tabular}{lrr}
\toprule
 & Count & Total \\
\midrule
$|\texttt{AOLIA}|$ & 1,525,524 & 93.4\% \\
$|\texttt{AOL17}|$ & 897,984 & 55.0\% \\
$|\texttt{AOL17}\setminus\texttt{AOLIA}|$ & 12,207 & 0.7 \% \\
$|\texttt{AOLIA}\setminus\texttt{AOL17}|$ & 639,747 & 39.2 \% \\
$|\texttt{AOLIA} \bigcup \texttt{AOL17}|$ & 1,537,731 & 94.2 \% \\
$|\texttt{AOLIA} \bigcap \texttt{AOL17}|$ & 885,777 & 54.3 \% \\
\bottomrule
\end{tabular}
\end{table}

\begin{figure}[b]
\centering
\includegraphics[scale=0.7]{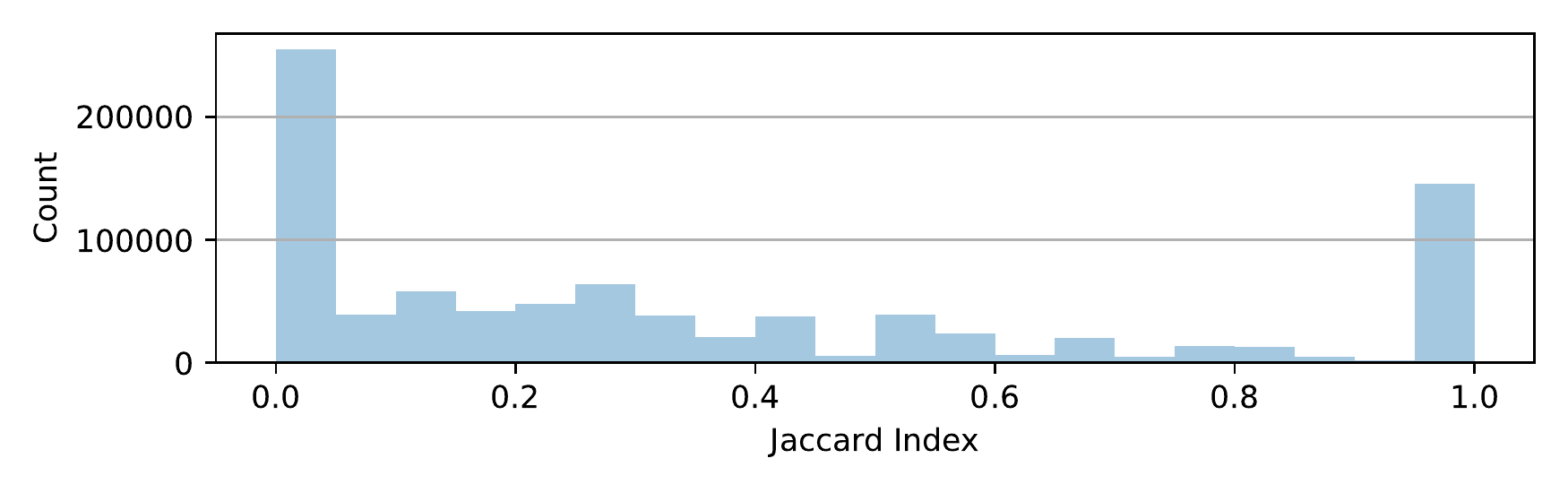}
\caption{Distribution of Jaccard similarities over title tokens from overlapping documents in \texttt{AOLIA} and \texttt{AOL17}.}
\label{fig:jaccard}
\end{figure}

We now dig into the characteristics of the 885,777 documents that overlap between the corpora. Fig.~\ref{fig:jaccard} presents the Jaccard similarity between the set of title tokens\footnote{Tokens considered are case-folded, alphanumeric strings separated by whitespace or punctuation.} present in each version of the document. Only 17\% of the titles have a perfect token overlap. Among these, 87\% are exact case-insensitive sequence matches, with typical differences being the replacement or addition of punctuation in the titles, but sometimes involves the repetition of words. Table~\ref{tab:title_diffs} shows such examples in rows 1--3.

The majority of documents (53\%) have at most a Jaccard index of 0.25, indicating low overlap, with 28\% having no overlap at all. In the cases when no overlap is present, semantically dissimilar content is often present, such as a placeholder document or a replacement with information about the web hosting provider. \sm{In semantically dissimilar cases, the queries that resulted in clicks for these documents are usually no longer relevant in \texttt{AOL17}.} For instance, the query ``indalo'' resulted in a click of the documented represented by \#4 in Table~\ref{tab:title_diffs}, which is a reasonable document in \texttt{AOLIA} but not in \texttt{AOL17}. In some cases, the content is semantically similar, such as in example \#6, and still likely a reasonable document for associated queries. Based on a manually-annotated random sample of 100 \sm{documents with Jaccard similarities of 0}, 23 appeared to be semantically-related (i.e., would likely satisfy similar information needs), while the remainder were not.

\begin{table}[t]
\centering
\caption{Examples of different titles found on pages between the \texttt{AOLIA} and \texttt{AOL17} corpora and their corresponding Jaccard index.}
\label{tab:title_diffs}
\resizebox{\textwidth}{!}{
\begin{tabular}{cllr}
\toprule
\# & \texttt{AOLIA} & \texttt{AOL17} & Jac. \\
\midrule
1 & Welcome To Atlanta Music Group ! & Welcome to Atlanta Music Group & 1.0 \\
2 & Vinopolis Wine Shop - Portland , Oregon & Vinopolis Wine Shop \textpipe~Portland , Oregon
 & 1.0 \\
3 & Mechanics Savings Bank & Mechanics Savings Bank - Mechanics Savings Bank & 1.0 \\
4 & Indalo Productions & InMotion Hosting & 0.0 \\
5 & Kennebec Valley Organization -- Home Page & \textit{(empty)} & 0.0 \\
6 & UK TV Guide & Homepage \textpipe~UKTV & 0.0 \\
7 & nutone // welcome & Nutone Records - Home \textpipe~Facebook & 0.2 \\
8 & Vedanta Press and Catalog & Books on Vedanta Philosophy & 0.1 \\
9 & Venning Graphic Utilities for blending images & Venning Graphic Utilities & 0.5 \\
10 & Steinway Musical Instruments , Inc . & Steinway Musical Instruments - Steinway \& Sons & 0.6 \\
\bottomrule
\end{tabular}}
\end{table}

\section{\edit{Reproduction and Replication}}\label{sec:repr}

\sm{The AOL Query Log has been used for training and evaluating numerous search tasks. In this section, we explore the effect of \texttt{AOLIA} on one such task: session-based personalisation. In this setting, a user's sequence of searches and clicks are broken into sessions (or tasks~\cite{Jones2008BeyondTS}). Within each session, the prior queries and clicks can act as additional context to help disambiguate the information needs of a query. For instance, if a user searches for ``cars'' followed by ``jaguar'', it is reasonable to tailor the results for the second query towards the luxury car brand rather than the animal. Although \sm{a few datasets are available for training and/or evaluating these systems (e.g., TREC Sessions~\cite{Kanoulas2011OverviewOT}), the AOL Query Log remains a popular (and often exclusive) choice for conducting these experiments.}}

\subsection{Methods}

We focus on three neural session-based personalisation techniques: M-NSRF~\cite{Ahmad2018MultiTaskLF}, M-MatchTensor~\cite{Ahmad2018MultiTaskLF}, and CARS~\cite{Ahmad2019ContextAD}. \edit{We select these methods because numerous recent works in the area use them as baselines} (e.g.,~\cite{Zhu2021ContrastiveLO,Cheng2021LongSS,Zhou2020EncodingHW,Qu2020ContextualRW}). \sm{All three models function as multi-task models, jointly learning to predict both document relevance and to predict the next query in the sequence. The models differ in the neural network architecture used to accomplish this. M-MatchTensor adapts the multi-task approach to the MatchTensor model~\cite{Jaech2017MatchTensorAD}, where MatchTensor builds a query-document similarity matrix between LSTM-encoded query and documet text and aggregates the results using CNN filters and max-pooling. M-NSRF encodes the query and document in separate bi-directional LSTM networks and combines them using a feed-forward layer to produce ranking scores. CARS builds upon M-NSRF by modeling the session interactions hierarchically using attention-based LSTM networks. For these three approaches, we use the authors' released code\footnote{\url{https://github.com/wasiahmad/context_attentive_ir}} with default parameters. In line with the code, the number of training iterations is tuned on dev data.}

In addition to the above \sm{task-specific} methods, we include three additional \sm{adhoc ranking} baselines to put the results in context. First, we use the Terrier~\cite{ounis2005terrier} BM25 implementation to re-rank the candidate documents with default BM25 parameters. This corresponds to the (unspecified) BM25 baseline conducted in~\cite{Ahmad2019ContextAD}. \sm{\io{Furthermore}, in light of recent findings in \textit{adhoc} retrieval, we also include two neural re-ranking baselines based on the T5 model~\cite{Nogueira2020DocumentRW,Raffel2020ExploringTL}.
Given that transferring relevance signals from one dataset/task to another using contextualised language models has generally \io{been} shown to be an effective technique (e.g.,~\cite{macavaney:emnlp2020-sledge}), we include a T5 ``transfer'' baseline. This version \edit{is} tuned on the MS MARCO dataset~\cite{Bajaj2016Msmarco}. We also use a ``tuned'' baseline, which continues model tuning on the AOL session data from the MS MARCO checkpoint (batch size 8, learning rate $5\times10^{-5}$). In line with \citet{Nogueira2020DocumentRW}, we simply train for a fixed number of batches without tuning this or other settings (here, 10,000 batches). For all three \sm{adhoc ranking} baselines, we use the PyTerrier~\cite{macdonald:cikm2021-pyterrier} implementation.}

\subsection{Experimental Settings}

We test the above \sm{six} systems on three settings. (1) Using both sessions and documents from the \texttt{AOL17} dataset. Here, we use the sessions constructed and provided by \citet{Ahmad2019ContextAD}. This reflects the original experimental setup and is therefore a study of \edit{reproducibility}. (2) Using the sessions from \texttt{AOL17} (provided by~\cite{Ahmad2019ContextAD}), but replacing the document titles with those from \texttt{AOLIA}. In the case where there is not a corresponding document in \texttt{AOLIA}, we leave the title blank \sm{(blanks present in 1.8\% of documents across training, dev, and test sets)}. Since the same sessions are used as in (1), these results are directly comparable and isolate the impact of the document text itself. (3) \edit{Using sessions from \texttt{AOLIA}}. In this setting, we \edit{re-create} sessions as described in~\cite{Ahmad2019ContextAD} \sm{(using same session elimination strategy and date ranges in each split)}, but using the full \texttt{AOLIA} corpus. Due to the higher coverage of documents, this \edit{results} in considerably more data and longer sessions \sm{across all three splits}. \sm{Table~\ref{tab:sessions} presents the characteristics of the sessions built by each dataset. \edit{Both (2) and (3) are replicability studies because  the experimental setting differs from the original paper (a different corpus is used than the original papers).}}

\sm{As is commonplace for the task, we use the document title for the document content. Given that many of the queries are navigational in nature, we also test a variant of each of the 3 above settings that also appends the tokenised URL to the title, which can allow models to distinguish between pages that have the same title content and to match queries that ask for a specific URL.}

\begin{table}[t]
\centering
\caption{Session search dataset characteristics provided by \texttt{AOL17} compared to \texttt{AOLIA}.}
\label{tab:sessions}
\scalebox{0.8}{
\setlength{\tabcolsep}{3pt}
\begin{tabular}{lrrrrrrrrr}
\toprule
& \multicolumn{3}{c}{Train} & \multicolumn{3}{c}{Dev} & \multicolumn{3}{c}{Test} \\
\cmidrule(lr){2-4}\cmidrule(lr){5-7}\cmidrule(lr){8-10}
& \texttt{AOL17} & \texttt{AOLIA} & \% & \texttt{AOL17} & \texttt{AOLIA} & \% & \texttt{AOL17} & \texttt{AOLIA} & \% \\
\midrule
\# Sessions & 219,748 & 311,877 & +42 \% & 34,090 & 49,522 & +45 \%  & 29,369 & 50,944 & +73 \% \\
\# Queries & 566,967 & 1,099,568 & +94 \% & 88,021 & 170,095 & +93 \% & 76,159 & 167,497 & +120 \% \\
Avg. Queries per Session & 2.58 & 3.53 & +37 \% & 2.58 & 3.43 & +33 \% & 2.59 & 3.29 & +27 \% \\
\bottomrule
\end{tabular}
}
\end{table}

\looseness -1 In all three settings, each of the supervised methods are each trained, tuned, and tested using the data from the \sm{corresponding} setting, while the unsupervised (BM25) and transfer (T5) baselines are simply run on the test set without tuning. We evaluate the results using MAP, MRR\footnote{Though this measure has been criticised~\cite{Fuhr2018SomeCM}, we report it to compare with past work.}, which are measures commonly-used for evaluation of this task. \edit{We calculate the measures using the \texttt{trec\_eval} implementation provided by \texttt{ir-measures}~\cite{macavaney:ecir2022-irm}. Note that this evaluation tool differs from the original work, which used their own implementation of the measures.} \citet{Qu2020ContextualRW} \edit{notes} that this can result in differences in measures due to tie-breaking behaviour. We conduct significance tests between all pairs of systems within each setting (paired t-test, $p<0.05$, with Bonferroni correction). \edit{We do not use a tool like \texttt{repro\_eval}~\cite{10.1007/978-3-030-72240-1_51} to compare our results with those from the original papers because the rankings provided from the original papers are not available.}

\subsection{Results}

\begin{table}[!h]
\centering
\caption{Comparison of personalised session search baselines using various versions of the datasets. Note that the \io{performances} of systems using \texttt{AOL17} and \texttt{AOLIA} sessions cannot be directly compared. \sm{The top results we measure for each setting are bold (i.e., not including results reported by others). \sm{Non-significant differences between pairs of runs within a setting are indicated with superscript letters (paired t-test, $p<0.05$, Bonferroni correction).}}}
\label{tab:main_results}
\begin{tabular}{cp{5cm}rrr}
\toprule
&Model & MAP & MRR & P@1 \\
\midrule

&\multicolumn{4}{c}{\textbf{(1) Sessions: \texttt{AOL17}, Documents: \texttt{AOL17}}} \\
a&BM25 (unsupervised) & 0.2457 & 0.2554 & 0.1454 \\
&\light{~- from~\cite{Ahmad2019ContextAD}}& \light{0.230\phantom{0}} & \light{0.206\phantom{0}} & \light{0.206\phantom{0}} \\
b&T5 (transfer) & 0.3553 & 0.3649 & 0.2242 \\
c&T5 (tuned)    & \bf 0.4538 & \bf 0.4640 & \bf 0.3001 \\
d&CARS          & $^e$0.4280 & $^e$0.4390 & $^e$0.2787 \\
&\light{~- from~\cite{Ahmad2019ContextAD}}& \light{0.531\phantom{0}} & \light{0.542\phantom{0}} & \light{0.391\phantom{0}} \\
&\light{~- from~\cite{Qu2020ContextualRW}}&\light{-} & \light{0.4538} & \light{0.2940} \\
e&M-MatchTensor & $^d$0.4335 & $^{df}$0.4444 & $^{df}$0.2830 \\
&\light{~- from~\cite{Ahmad2019ContextAD}}& \light{0.505\phantom{0}} & \light{0.518\phantom{0}} & \light{0.368\phantom{0}} \\
f&M-NSRF        & 0.4410 & $^e$0.4521 & $^e$0.2904 \\
&\light{~- from~\cite{Ahmad2019ContextAD}}& \light{0.491\phantom{0}} & \light{0.502\phantom{0}} & \light{0.391\phantom{0}} \\

\midrule
&\multicolumn{4}{c}{\textbf{(2) Sessions: \texttt{AOL17}, Documents: \texttt{AOLIA}}} \\
a&BM25 (unsupervised) & 0.2942 & 0.3044 & 0.1914 \\
b&T5 (transfer) & 0.4228 & 0.4337 & 0.3021 \\
c&T5 (tuned)    & \bf 0.5115 & \bf 0.5223 & \bf 0.3745 \\
d&CARS          & 0.4998 & 0.5107 & 0.3630 \\
e&M-MatchTensor &$^f$0.4848 &$^f$0.4961 &$^f$0.3493 \\
f&M-NSRF        &$^e$0.4911 &$^e$0.5023 &$^e$0.3495 \\

\midrule
\midrule
&\multicolumn{4}{c}{\textbf{(3) Sessions: \texttt{AOLIA}, Documents: \texttt{AOLIA}}} \\
a&BM25 (unsupervised) & 0.2413 & 0.2413 & 0.1462 \\
b&T5 (transfer) & 0.3620 & 0.3620 & 0.2260 \\
c&T5 (tuned)    &\bf0.4171 &\bf0.4171 &\bf0.2650 \\
d&CARS          & 0.3784 & 0.3784 & 0.2294 \\
e&M-MatchTensor & 0.3572 & 0.3572 & 0.2133 \\
f&M-NSRF        & 0.4009 & 0.4009 & 0.2534 \\
\bottomrule
\end{tabular}
\end{table}

\sm{Table~\ref{tab:main_results} presents the results for the three settings when using the document title as its contents. In Setting (1), where we use the same data and code as~\cite{Ahmad2019ContextAD}, we \edit{are} unable to reproduce the performance reported by \citet{Ahmad2019ContextAD}. However, we note that the performance for CARS is not far from the results reported by \citet{Qu2020ContextualRW}, who \edit{report} that the discrepancies with the original work \edit{are} due to using the \texttt{trec\_eval} measure implementation. Setting (2), where we use \texttt{AOLIA} document titles, yields performances closer to those reported by \citet{Ahmad2019ContextAD}, with the CARS model outperforming M-MatchTensor and M-NSRF. Finally, when using both sessions and documents from \texttt{AOLIA} (3), M-NSRF outperforms the other two session search methods. In short, we validate the findings of \citet{Ahmad2019ContextAD} that CARS significantly outperforms M-MatchTensor and M-NSRF, but only when using \texttt{AOL17} sessions and \texttt{AOLIA} documents; in the other two settings, we draw the conclusion that the (simpler) M-NSRF model significantly outperforms the other two approaches.}

\sm{Across all three settings, however, the adhoc (i.e., session unaware) tuned T5 model outperforms all other methods. We acknowledge that comparing T5 with CARS, M-MatchTensor, and M-NSRF is not a completely fair comparison; the T5 model benefits from a much larger model and extensive pre-training, while CARS, M-MatchTensor, and M-NSRF benefit from access to past queries and clicks within the session. Techniques for adapting contextualised language models like T5 for session search have been explored in~\cite{Qu2020ContextualRW}, though in pilot studies we had difficulty training effective models using the released code. We also note that although T5 benefits from tuning on the target domain, it can still perform reasonably well---especially in Setting (3), where \io{the} \texttt{AOLIA} documents and sessions \edit{are} used.}

\begin{table}[!ht]
\centering
\caption{Results when including the tokenised URL in addition to the title. The $\Delta$ column indicates the improvements compared to the results without the URL (Table~\ref{tab:main_results}). \sm{Non-significant differences between pairs of runs within a setting are indicated with superscript letters (paired t-test, $p<0.05$, Bonferroni correction).}}
\label{tab:url_results}
\scalebox{1.0}{
\setlength{\tabcolsep}{2pt}
\begin{tabular}{clrrrrrr}
\toprule
&Model & MAP & $\Delta$& MRR & $\Delta$& P@1 & $\Delta$\\
\midrule

&\multicolumn{7}{c}{\textbf{(1) Sessions: \texttt{AOL17}, Documents: \texttt{AOL17}}} \\
a&BM25 (unsupervised) & 0.3204 & +0.0747 & 0.3314 & +0.0760 & 0.1991 & +0.0537 \\
b&T5 (transfer) & 0.5023 & +0.1470 & 0.5135 & +0.1486 & 0.3572 & +0.1330 \\
c&T5 (tuned)    & \bf 0.7074 & \bf +0.2536 & \bf 0.7190 & \bf +0.2550 & \bf 0.6201 & \bf +0.3200 \\
d&CARS          & $^f$0.6530 & +0.2250 & $^f$0.6643 & +0.2253 & $^f$0.5493 & +0.2706 \\
e&M-MatchTensor & 0.6756 & +0.2421 & 0.6871 & +0.2427 & 0.5784 & +0.2954 \\
f&M-NSRF        & $^d$0.6634 & +0.2224 & $^d$0.6745 & +0.2224 & $^d$0.5602 & +0.2698 \\

\midrule
&\multicolumn{7}{c}{\textbf{(2) Sessions: \texttt{AOL17}, Documents: \texttt{AOLIA}}} \\
a&BM25 (unsupervised) & 0.3484 & +0.0542 & 0.3591 & +0.0547 & 0.2360 & +0.0446 \\
b&T5 (transfer) & 0.5400 & +0.1172 & 0.5514 & +0.1177 & 0.3959 & +0.0938 \\
c&T5 (tuned)    &\bf0.7071 &\bf+0.1956 &\bf0.7183 &\bf+0.1960 &\bf0.6153 &\bf+0.2408 \\
d&CARS          & 0.6665 & +0.1667 & 0.6774 & +0.1667 & $^e$0.5660 & +0.2030 \\
e&M-MatchTensor & $^f$0.6538 & +0.1690 & $^f$0.6654 & +0.1693 & $^{df}$0.5569 & +0.2076 \\
f&M-NSRF        & $^e$0.6520 & +0.1609 & $^e$0.6632 & +0.1609 & $^e$0.5501 & +0.2006 \\

\midrule
\midrule
&\multicolumn{7}{c}{\textbf{(3) Sessions: \texttt{AOLIA}, Documents: \texttt{AOLIA}}} \\
a&BM25 (unsupervised) &   0.2997 &   +0.0584 &   0.2997 &   +0.0584 &   0.1790 &   +0.0328 \\
b&T5 (transfer)       &   0.4260 &   +0.0640 &   0.4260 &   +0.0640 &   0.2693 &   +0.0433 \\
c&T5 (tuned)          &\bf0.5679 & +0.1508 &\bf0.5679 & +0.1508 &\bf0.4418 & +0.1768  \\
d&CARS                &   0.5360 &   +0.1576 &   0.5360 &   +0.1576 &   0.4082 &   +0.1788 \\
e&M-MatchTensor       &   0.5458 &\bf+0.1886 &   0.5458 &\bf+0.1886 &$^f$0.4297 &\bf+0.2164 \\
f&M-NSRF              & 0.5575 &   +0.1566 & 0.5575 &   +0.1566 & $^e$0.4336 &   +0.1802 \\

\bottomrule
\end{tabular}
}
\end{table}

\sm{Table~\ref{tab:url_results} presents the results in each setting when appending the URL to the document text. We observe that in every case, this additional feature improves ranking effectiveness, sometimes by a considerable margin (up to +0.25 in MAP, +0.26 in MRR, and +0.32 in P@1). These findings underscore the importance of including this signal when queries are often navigational in nature.}

\sm{Overall, we find that these experiments provide further evidence \io{that} \texttt{AOLIA} is well-constructed and useful. Between Settings (1) and (2), we see a consistent boost in ranking effectiveness across several models. Since the only thing \edit{we change} between these settings \edit{is} the document text, it suggests that the texts in 
\texttt{AOLIA} are more in line with the preferences of the users. Our experiments using T5 and URL features suggest that more care should be taken in future session-based search studies to construct evaluation data that focus on information needs that are less navigational in nature, as these can be addressed simply using established adhoc approaches and navigational signals.}

\section{Limitations}\label{sec:limit}

One limitation of using The Internet Archive are actions that the organisation takes in response to copyright claims made against archived \io{content}, which can effectively remove documents from the archive.\footnote{See their official copyright policy here: \url{https://archive.org/about/terms.php}} \edit{Over} an approximately one-month window, 51 documents originally present in a prior version of \texttt{AOLIA} were no longer available on the Internet Archive, presumably due to this policy. Though this has the potential for knock-on effects downstream, prior work~\cite{McCreadie2012OnBA} indicates that it will likely have little effect. Specifically, \citet{McCreadie2012OnBA} \edit{find} that the effect of a far greater proportion of documents being deleted from the TREC Microblog 2011 corpus had \io{a minimal effect} on system evaluation, \io{hence} we expect the same to be true for \texttt{AOLIA}. \io{Furthermore}, when compared with the vast \edit{proportion} of documents missing in more recent versions of the AOL corpus (e.g., \texttt{AOL17} is missing 746,843 documents), the potential for several hundred removed documents per year seems preferable. Nevertheless, future work studying these effects may be warranted.

\sm{Although \texttt{AOLIA} improves the coverage and contents of documents that appear in the log, it does not attempt to fill in other documents that may have appeared in the corpus. In this way, it does not reflect a realistic sample of documents that likely existed in the entire AOL corpus at the time; it is more akin to the MS MARCO v1 passage corpus~\cite{Bajaj2016Msmarco} (which only includes passages that were presented to annotators) than to corpora like ClueWeb (which includes documents scraped using typical web crawling techniques). As noted in Section~\ref{sec:back}, \edit{available web crawls would provide} a less accurate picture of the documents as they appeared to the AOL users. \edit{Therefore,} despite this limitation, we believe \texttt{AOLIA} is still valuable in many practical experimental settings.}

\section{Conclusions}\label{sec:conc}

\sm{In this work, we studied problems surrounding reproducibility of the AOL Query Log's corpus. We started by carefully constructing a new version of the corpus that better reflects the documents as they appeared to the users at the time of the log. We found that our approach increases the coverage of documents in the log considerably, when compared to a version that scraped documents eleven years after the log. We further found that the contents of documents are prone to considerable change over time, with the majority of document titles having very low token overlap between versions. When reproducing prior results for session search, we find that our new corpus improves the effectiveness across a variety of models (likely attributable to more realistic documents), and brings benchmarks based off the AOL Query Log more in line with adhoc ranking methods. We \edit{made} access to our new version of the AOL corpus easily available to assist in future reproducibility efforts.}

\vspace{1em}
\noindent\textbf{Acknowledgments.} We acknowledge EPSRC grant EP/R018634/1: Closed-Loop Data Science for Complex, Computationally- \& Data-Intensive Analytics. \sm{We thank \citeauthor{Ahmad2019ContextAD} for sharing the data and code that facilitated this study.}

\bibliographystyle{splncs04nat}
\bibliography{blblio.bib}
\end{document}